\documentstyle[aps,epsf]{revtex}

\newcommand \beq {\begin{equation}}
\newcommand \eeq {\end{equation}}
\newcommand \boldsigma {\mbox{\boldmath $\sigma$}}
\newcommand \boldxi {\mbox{\boldmath $\xi$}}
\newcommand \bolds {\mbox{\boldmath $s$}}
\newcommand \rra {\rangle \rangle}
\newcommand \lla {\langle \langle}
\newcommand \bay {\begin{array}}
\newcommand \eay {\end{array}}

\begin{document}

\title{Multifractal analysis of perceptron learning with errors}
\author{ M. Weigt  \thanks{martin.weigt@physik.uni-magdeburg.de}  \\
      {\small{\it Institut f\"ur Theoretische Physik,
      Otto-von-Guericke-Universit\"at Magdeburg}}\\
      {\small{\it PSF 4120,
      39016 Magdeburg, Germany}}}
 
\date{\small July, 15, 1997}
\maketitle
 
\begin{abstract}
Random input patterns induce a partition of the coupling space of a 
perceptron into
cells labeled by their output sequences. Learning some data with
a maximal error rate leads to clusters of neighboring cells. By analyzing the
internal structure of these clusters with the formalism of multifractals, we
can handle different storage and generalization tasks for lazy students and
absent--minded teachers within one unified approach. The results also allow
some conclusions on the spatial distribution of cells. \\[0.5cm]
\end{abstract}

\section{Introduction}

Artificial neural networks show considerable information processing 
capabilities, see
e.g. \cite{KHP}. One of the most important tasks is {\em classification} 
of data
according to an initially unknown rule. Considering a set of 
$p=\gamma N$ input patterns $\boldxi^\mu\in I\!\!R^N, \mu=1,...,p$, 
there are $2^p$ possible binary functions
$\boldxi^\mu \mapsto \sigma^\mu = \pm 1$. Some of them are linearly
separable and can be realized by a simple perceptron
\beq
\label{perc}
\sigma = \mbox{sgn} ( {\bf J}\cdot\boldxi ) = \mbox{sgn} ( \sum_i J_i\xi_i )
\eeq
where ${\bf J}\in I\!\!R^N$ is called the {\em coupling vector}. 
Due to the scaling invariance
of (\ref{perc}) this vector can be restricted by the 
{\em spherical constraint} ${\bf J}\cdot{\bf J}=N$.
The direction of ${\bf J}$ fixes the actual form of the classification.

Not all coupling vectors define different functions of the $p$ 
input patterns. According to their possible output sequences 
$\boldsigma = \{ \sigma^\mu|\mu=1,...,p \}$ 
we can group them together into at most $2^p$ {\em cells}
\beq
\label{celldef}
C( \boldsigma ) = 
\{ {\bf J}|\;  \sigma^\mu = \mbox{sgn} ( {\bf J}\cdot\boldxi^\mu ) 
\;\forall \mu \}\;.
\eeq
These cells form a partition of the coupling space whose structure 
contains important information on the performance of the perceptron
in various supervised learning problems.

The use of statistical mechanics in the study of the coupling space for 
large $N$ was
initiated by Gardner \cite{gardner} for random input--output relations. 
Derrida et al.
\cite{DGP} suggested calculating the {\em cell size distribution}, which 
could be done
only two years later when Monasson and O'Kane \cite{MoKa} introduced 
a modification of the standard replica trick in connection with multifractal
techniques.
Now there are several applications for perceptrons \cite{EW,WE,BO,BvdB} and
multilayer networks \cite{MZ,CMZ}.

All these calculations consider the case where a uniquely determined output
is perfectly learned by the student network. However, there is often no need 
or no possibility of perfect learning a special classification, or in real 
applications only noisy output data are available. Introducing an
error rate corresponds to collecting several cells (\ref{celldef})
into clusters. In the present paper we use the multifractal approach to 
characterize the coupling space structure of the output 
representations in these clusters. This analysis allows us
to observe various storage and generalization problems within one approach.
We include both the case of a student who perfectly learns some
incorrect data (e.g. generalization with output noise)
as well as the case of a student who tries to learn a well--defined task only 
with a certain error rate (e.g. storage with minimal error above the storage
capacity).

The outline of the paper is as follows. In Sec. II we present the multifractal 
formalism for neural networks. Sec. III
contains the general calculations for the internal representations of the cell
clusters. In Sec. IV and V the most interesting cases are analyzed in detail,
i.e. the storage and the
generalization problems with noise. In Sec. VI we briefly comment
on the spatial 
distribution of the cells. A summary is given in the final section.

\section{The multifractal formalism}

Due to the geometrical nature of our problem a multifractal method
is the appropriate one.
In this section we introduce the multifractal formalism as applied
to perceptrons. In order to  clarify the notation we review some 
results obtained in \cite{EW,WE} for spherical couplings without going into
the subtleties of the approach.

We choose $p = \gamma N$ input patterns
$\boldxi^\mu \in \{-1,1\}^N,\, \mu = 1,\ldots ,p$, with
entries randomly drawn from the distribution 
$p(\xi_i^\mu) = 1/2\; \delta(\xi_i^\mu+1) + 1/2\; \delta(\xi_i^\mu-1)$.
The hyperplane perpendicular to each $\boldxi^\mu$ cuts the coupling
space into two parts. The patterns
therefore generate a random partition of the coupling space into cells
defined by (\ref{celldef}) and labeled by their output sequences $\boldsigma$.
The relative cell size
$P(\boldsigma)=V(\boldsigma ) /$
$\sum_{{\mbox{\boldmath $\tau$}}}V({\mbox{\boldmath $\tau$} })$
describes the probability of generating the output ${\boldsigma}$ for
a given input sequence ${\mbox{\boldmath $\xi$}}^\mu$ with a coupling vector
${\bf J}$ chosen at random from a uniform distribution over the whole 
coupling sphere. In the thermodynamic limit they are expected to scale
exponentially with $N$, consequently we characterize the cell sizes by the 
{\it crowding index} 
$\alpha( \boldsigma )$ defined by
\beq
P( \boldsigma ) = 2^{-N\alpha( \boldsigma)} \;.
\eeq
The storage and generalization properties of the perceptron are coded in the 
{\em distribution of cell sizes} defined by 
\beq
f(\alpha)=\lim_{N\to\infty} \frac{1}{N} \log_2 
    \sum_{\boldsigma}\delta(\alpha-\alpha(\boldsigma)) \;.
\eeq
In the language of multifractals this quantity is called the 
{\em multifractal spectrum}.
To calculate it within the framework of statistical mechanics one
uses the formal analogy of $f(\alpha)$ with the micro--canonical entropy of
the spin system $\boldsigma$ with Hamiltonian $N\alpha(\boldsigma)$. 
It can hence be determined from the corresponding `free energy'
\beq\label{selfaver}
\tau ({p}) = - \lim_{N\to\infty} \frac{1}{N}  \log_2 \sum_{\boldsigma}
            2^{-{p}N\alpha(\boldsigma)} 
= - \lim_{N\to\infty} \frac{1}{N} \log_2 \sum_{\boldsigma}
       P^{p} (\boldsigma )  
\eeq
via Legendre-transformation with respect to the ``inverse temperature'' ${p}$
\beq\label{legendre}
f(\alpha) = \min_{{p}} [\alpha {p} - \tau({p})]\;.
\eeq
In the multifractal terminology $\tau({p})$ is called the {\it mass exponent}.

To explicitly calculate this quantity for the perceptron we start with the 
definition of the cell size
\beq
P(\boldsigma) = \int d\mu({\bf J}) \prod_{\mu =1}^p 
\theta( \frac{1}{\sqrt{N}}
\sigma^\mu {\bf J}\cdot \boldxi^\mu)
\label{size}
\eeq
using the Heaviside step function $\theta(x)$. The integral measure 
\beq
d\mu({\bf J}) = \prod_i \frac{dJ_i}{\sqrt{2\pi e}} \; \delta(N-{\bf J}^2) .
\eeq
ensures both the spherical constraint for the coupling vectors as well as
the total normalization $\sum_{\boldsigma} P(\boldsigma)=1$. 

In the thermodynamic limit $N\to\infty$ we expect both $\tau$ and $f$ 
to become self-averaging,
and we can therefore calculate the mass exponent (\ref{selfaver}) by
using the replica trick introducing $n$ identical
replicas numbered $a=1,\ldots,n$ to perform the average over the quenched 
patterns.
Moreover, we introduce a second
replica index $\alpha = 1,\ldots,{p}$ in order to represent the ${p}$-th power
of $P$ in (\ref{selfaver}). Using an integral representation 
for the  Heaviside function we arrive at a replicated partition function
given by
\begin{eqnarray}
Z_n & = &  \lla Z^n \rra \nonumber\\
  & = & \lla  \sum_{\{ \sigma_\mu^a \} } \int\prod_{a,\alpha}
d\mu ( {\bf J}_a^\alpha) \; \prod_{\mu,a,\alpha}\theta(
\frac{\sigma_\mu^a}{\sqrt{N}}{\bf J}_a^\alpha \cdot\boldxi^{\mu}) 
  \rra \nonumber \\
 &=&\lla\sum_{\{ \sigma_\mu^a \} } \int \prod_{a,\alpha}
d\mu ( {\bf J}_a^\alpha ) \int_0^\infty \prod_{\mu,a,\alpha} 
\frac{d\lambda_\mu^{ a,\alpha}}
{\sqrt{2\pi}} \int \prod_{\mu,a,\alpha} \frac{dx_\mu^{
a,\alpha}}{\sqrt{2\pi}} \nonumber \\
 & & \exp\left\{ i \sum_{\mu,a,\alpha} 
 x_\mu^{a,\alpha} ( \lambda_\mu^{a,\alpha} - \frac{\sigma_\mu^a}{\sqrt{N}} 
  {\bf J}_a^\alpha \cdot\boldxi^\mu ) \right\} \rra \;.
\label{massgen}
\end{eqnarray}
As usual, the average $\lla\cdot\rra$ over the quenched 
patterns can be performed, and the overlaps
\beq
P_{a,b}^{\alpha,\beta} = \frac{1}{N} {\bf J}_a^\alpha\cdot{\bf J}_b^\beta
\label{overlap}
\eeq
are introduced as order parameters. The spherical constraint 
restricts the diagonal elements of this matrix to one.

It is important to note that the output sequences $\{ \sigma_\mu^a \}$ carry 
only one replica index. The typical overlap of two coupling vectors within
one cell (same output sequence $\{ \sigma_\mu^a \}$) will hence 
in general be different
from the typical overlap between two coupling vectors belonging to different
cells (different output sequence $\{ \sigma_\mu^a \}$).
Therefore we have to introduce already within the {\em replica symmetric 
approximation} two different overlap values:
\beq
P_{a,b}^{\alpha,\beta} = \left\{
\bay{ll}
1 & \mbox{if} \;\; (a,\alpha)=(b,\beta) \\
P & \mbox{if} \;\; a=b,\;\alpha\neq\beta \\
P_0 & \mbox{if} \;\; a\neq b
\eay \right. \;.
\label{QRS}
\eeq
In accordance with the above discussion $P$ then denotes the typical
overlap {\em within
one cell}, whereas $P_0$ denotes the overlap {\em between different cells}. 

Plugging this RS ansatz into (\ref{massgen}) one
realizes that  $P_0 = 0$ always solves the saddle point equations
for $P_0$. This has an obvious 
physical interpretation: Due to the symmetry of (\ref{perc}) and therefore of 
the crowding index under the transformation 
$({\bf J}, \boldsigma ) \leftrightarrow (-{\bf J}, -\boldsigma )$ every cell 
has a ``mirror cell'' of same size and shape on the ``opposite side'' of the 
coupling  space. $P_0=0$ simply reflects this symmetry. 

Finally we obtain the mass exponent 
\beq
\tau({p}) = - \frac{1}{\log 2}  \;\mbox{extr}_{P}\left[ \frac{1}{2}
  \log (1+({p}-1)P) - 
\frac{{p}-1}{2} \log ( 1-P ) - \gamma \log \; 2 \int Dt H^{p}\left( 
\sqrt{ \frac{P}{1-P} } t \right)\right]
\label{globalrsmass}
\eeq
where we introduced the abbreviations $Dt = dt\; \exp(-t^2/2) / \sqrt{2\pi}$ 
for the Gaussian measure and $H(x) = \int_x^{\infty} Dt$.
The order parameter $P$ is 
self-consistently determined by the saddle point equation
\beq
\frac{P}{1+({p}-1)P} = \frac{\gamma}{2\pi} 
\frac{\int Dt H^{{p}-2}\left( \sqrt{ \frac{P}{1-P} } t \right)
\exp \left\{ -\frac{P}{1-P} t^2 \right\} }
{\int Dt H^{{p}}\left( \sqrt{ \frac{P}{1-P} } t \right) } \;.
\label{saddle1}
\eeq
This equation can only be solved numerically, the results
are shown in fig. 1. 

\begin{figure}[htb]
 \epsfysize=10cm
      \epsffile{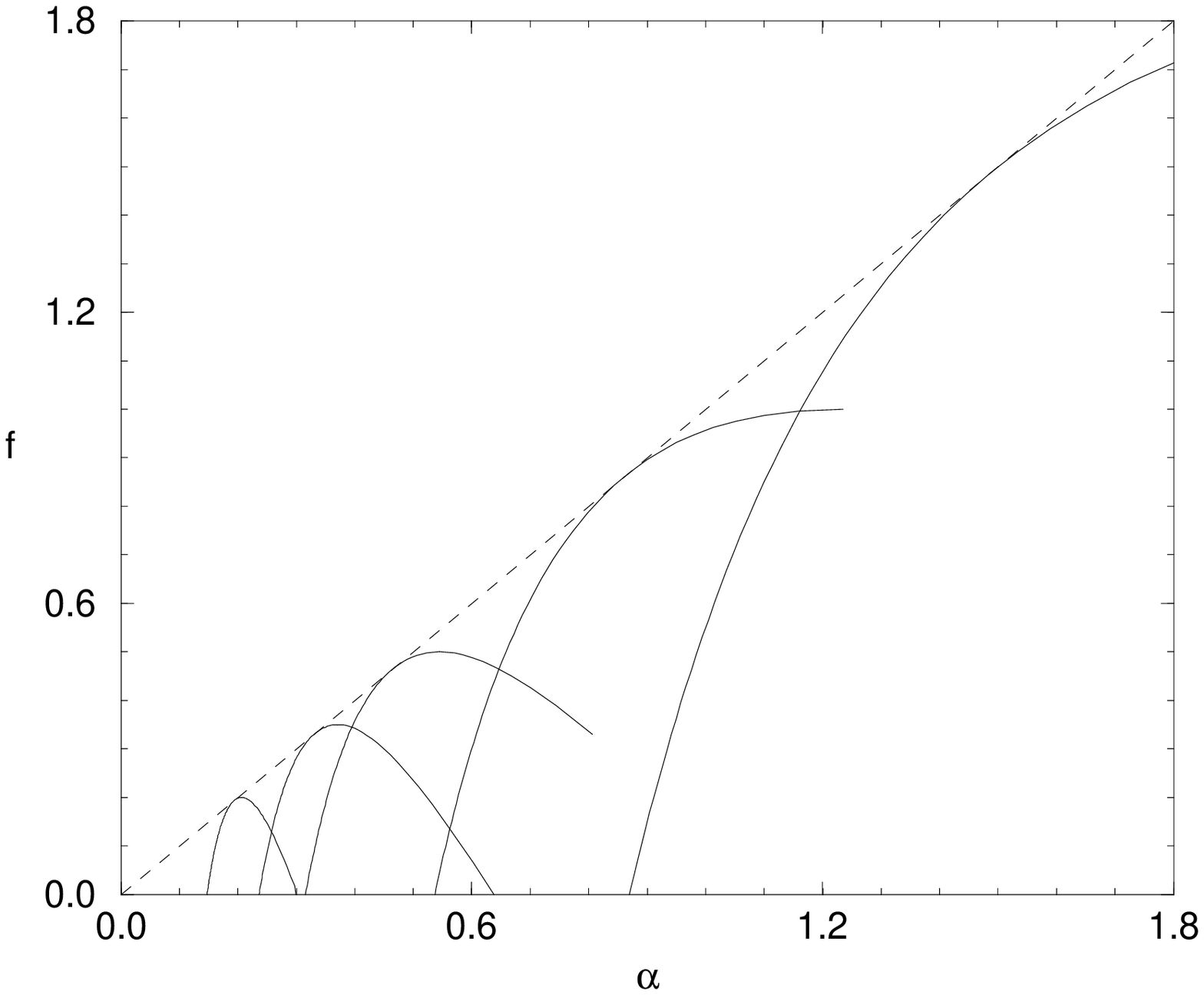}
\caption
{Multifractal spectrum $f(\alpha)$ characterizing the cell structure of
the coupling space of the spherical perceptron for various values
of the loading parameter $\gamma=0.2,0.35,0.5,1.0,2.0$ (from left to right).}
\end{figure}

The total number of cells is given by
\beq
{\cal{N}}=\int_0^{\infty} d\alpha 2^{Nf(\alpha)}
\eeq
and is therefore exponentially dominated by cells of size
$\alpha_0(\gamma)= \mbox{argmax}(f(\alpha))$. Because of
$\frac{df}{d\alpha} = p$ this point is reached at $p=0$.
 The random choice of any output
sequence will hence lead with probability one to a cell of size
$\alpha_0(\gamma)$, and $2^{-N\alpha_0(\gamma)}$ is found to be the Gardner 
volume. From $\alpha_0(\gamma\to 2)\to\infty$ we find the storage capacity 
to be $\gamma_c=2$ as in \cite{gardner}. For
$\gamma<2$ the problem of storing $\gamma N$ random input--output pairs
is realizable with probability one. So we have ${\cal{N}}=2^{\gamma N-o(N)}$
and therefore $f(\alpha_0(\gamma))=\gamma$ in the thermodynamic limit.

Although the cells with volume $\alpha_0$ are the most frequent ones, their
joint contribution to the total {\em volume} of the sphere is negligible.
Since 
\beq
1=\sum_{\boldsigma} P(\boldsigma)=\int_0^{\infty} d\alpha
    2^{N[f(\alpha)-\alpha]}
\eeq
a saddle point argument reveals that the cells with size $\alpha_1(\gamma)$
defined by $\frac{df}{d\alpha}(\alpha_1)=1$ dominate the volume. 
This point is given by ${p}=1$. Cells of larger size are
too rare, those more frequent are too small to compete. Consequently a
randomly chosen {\em coupling vector} ${\bf J}$ will with probability one
belong to a cell of size $\alpha_1$. By the definition (\ref{celldef}) 
of the cells
all other couplings of this cell will give the same output for all patterns
 $\boldxi^{\mu}$. Therefore $2^{-N\alpha_1(\gamma)}$ is nothing but
the volume of the version space of a teacher perceptron chosen at random
from a  uniform probability distribution 
on the sphere of possible perceptrons. From it (or equivalently
from $P({p}=1,\gamma))$ one can determine the generalization error as a
function of the training set size $\gamma$, thus reproducing the results of
\cite{GT}. 

\section{Internal cell structure of clusters for noisy output data}

In order to include noisy output data we have to slightly modify this 
procedure. As before, we consider
a randomly drawn set of input patterns $\{\boldxi^\mu;\mu=1,...,\gamma N\}$ 
as quenched disorder. The global cell distribution consequently equals the
one in the previous section.

Now we take any output sequence $\bolds\in \{-1,1\}^{\gamma N}$,
demanding it to be learned with an
error rate $\delta \in (0,0.5)$. $\delta=0$ corresponds to the noise--free 
case, $\delta=0.5$ to
outputs which are totally uncorrelated to the original pattern $\bolds$. 
The realized output
$\boldsigma\in \{-1,1\}^{\gamma N}$ has an overlap
\beq
\label{error}
\frac{1}{\gamma N}\sum_{\mu=1}^{\gamma N}\sigma^\mu s^\mu = 1-2\delta 
\eeq
with $\bolds$. The set of all cells $C(\boldsigma)$ with this
output overlap forms a cluster. It is the internal structure of the cluster
which we will analyze, i.e. we calculate the internal cell spectrum
of the cluster. The restricted partition function can be written as
\beq
\label{pf}
Z(\bolds ,\delta)=\sum_{\boldsigma} 
\delta(\frac{1}{\gamma N}\sum_{\mu=1}^{\gamma N}\sigma^\mu s^\mu -1+2\delta)
P^q(\boldsigma) \;,
\eeq
where the relative volume $P(\boldsigma)$ is defined by (\ref{size}).
In the special case of a randomly drawn sequence $\bolds$ this quantity
is closely related to the partition function considered in \cite{RS} where
a Gibbs measure of the error rate was introduced instead of the 
$\delta$-- function in (\ref{pf}).
Being self--averaging $Z(\bolds,\delta)$ does not depend on $\bolds$ itself,  
but only on the size $\alpha(\bolds)$ of the central cell. It can therefore 
be characterized by the real number $p$ with 
$\alpha(\bolds)=\alpha_p, \frac{df}{d\alpha}(\alpha_p)=p,$ in the global
spectrum. The mass exponent of the cluster is thus given by
\beq
\label{massexp}
\tau(q|{p},\delta) = -\lim_{N\to\infty}\frac{1}{N} 
         \lla \frac{\sum_{\bolds} P^{{p}}(\bolds) 
         \log_2 Z(\bolds ,\delta) }{
                  \sum_{\bolds} P^{{p}}(\bolds) } \rra \;.
\eeq
This is in complete analogy to the standard calculation of canonical 
expectation values in statistical
mechanics. A very similar method was introduced in \cite{FP} in order 
to characterize metastable states
in spherical $p$--spin glasses. In that case one system was thermalized 
in an equilibrium state, whereas a
second one was restricted to have a certain overlap with the first one -- 
which is analogous to our output sequences $\bolds$ and $\boldsigma$.

The internal spectrum $f(\alpha|{p},\delta)$ of the cluster can again
be calculated by a Legendre transformation with respect to the 
`inverse temperature' $q$, cf. (\ref{legendre}).

Before explicitly performing the technical part of the analysis we want to
clarify the question which problems can be solved within our approach.
Clearly, the value of $p$ fixes the original learning task without noise, which
corresponds to perfect learning of the output sequence $\bolds$. 
As already discussed in Sec. II, ${p}=0,1$ are of particular importance for
storage and generalization problems. 
Now, $q=0$ describes the most frequent cell within the cluster. If we take
any random output string $\boldsigma$ having overlap $1-2\delta$ with 
$\bolds$, we will arrive with probability one in a cell of size 
$\alpha(q=0|{p},\delta)$. This point corresponds therefore to a student
who perfectly learns one particular incorrect output sequence. For the 
generalization problem, ${p}=1$, it gives the behavior in the presence
of output noise. 

On the other hand, $q=1$ characterizes the volume--dominating
cells of the cluster, the total crowding index of the cluster is given by
\beq
\alpha_{cl}({p},\delta) = \alpha(q=1|{p},\delta)-
f(\alpha(q=1|{p},\delta)|{p},\delta)\;
\eeq
It describes the volume of the version space of a lazy student 
who is satisfied whenever he finds a coupling vector producing errors
with maximal rate $\delta$.

From the spectra for different ${p}$ but fixed $\delta$ we can 
get some information on the spatial
distribution of the cells -- whether there are more large/small cells in 
the environment of another large/small cell. This can be read off the
${p}$--dependence of the internal cluster spectrum for one value of $\delta$.

In order to answer all these questions we have to calculate the 
mass exponent (\ref{massexp}).
We need to introduce four replications as representation of:\\
(i) the logarithm of the partition function: $a=1,...,n$,\\
(ii) the power $q$ in the partition function: $\alpha=1,...,q$,\\
(iii) the fraction in the average over all ${p}$--cells: $k=1,...,m$,\\
(iv) the power ${p}$ in the average over all ${p}$--cells: 
  $\kappa=1,...,{p}$.\\
The replicated and averaged partition function consequently reads
\beq
\label{partition}
Z_{m,n} = \lla \sum_{\bolds_k,\boldsigma_a} \int\prod_{k,\kappa} 
d\mu({\bf K}_k^\kappa) 
              \prod_{\mu,k,\kappa}\Theta(\frac{s^{\mu}_{k}}{\sqrt{N}}
{\bf K}_k^{\kappa}\cdot\boldxi^\mu)
               \int\prod_{a,\alpha} d\mu({\bf J}_a^\alpha) \prod_{\mu,a,\alpha}
               \Theta(\frac{\sigma^{\mu}_{a}}{\sqrt{N}}{\bf J}_a^{\alpha}
\cdot\boldxi^\mu)
               \;\prod_a \delta(\frac{1}{\gamma N}
                   \bolds_1\cdot\boldsigma_a-1+2\delta)
              \rra\;.
\eeq
The coupling vectors ${\bf K}_k^\kappa$ are elements of the ${p}$--cells,
${\bf K}_1^\kappa$ of the central cell of the cluster. ${\bf J}_a^\alpha$
lies within the cluster cells.
Using this, the mass exponent can be determined from the replica trick
\beq
\tau(q|{p},\delta)=-\lim_{N\to \infty}\frac{1}{N\ln 2}\lim_{m,n\rightarrow 0} 
\partial_n Z_{m,n} \;.
\eeq
The calculation of $Z_{m,n}$ widely follows standard routes and uses 
the order parameters
\begin{eqnarray}
P_{k,l}^{\kappa,\lambda} & = & \frac{1}{N} 
{\bf K}_k^{\kappa}\cdot{\bf K}_l^{\lambda} \;,\;\;
     \forall k,l=1,...,m;\kappa,\lambda=1,...,{p}\nonumber \\
Q_{a,b}^{\alpha,\beta} & = & \frac{1}{N} {\bf J}_a^{\alpha}
\cdot{\bf J}_b^{\beta} \;,\;\;\;\;
     \forall a,b=1,...,n;\alpha,\beta=1,...,q \\
R_{k,a}^{\kappa,\alpha} & = & \frac{1}{N} {\bf K}_k^{\kappa}
\cdot{\bf J}_a^{\alpha} \nonumber
\end{eqnarray}
for the overlaps of coupling vectors from ${p}$--cells and from 
$q$--cells of the cluster. The diagonal
elements of the matrices ${\bf Q}$ and ${\bf P}$ are restricted 
to one by the spherical constraint.
This leads after standard manipulations to
\begin{eqnarray}
\label{Z_mn}
Z_{m,n} & = & \int\prod_{(k,\kappa)<(l,\lambda)} dP_{k,l}^{\kappa,\lambda}
            \int\prod_{(a,\alpha)<(b,\beta)} dQ_{a,b}^{\alpha,\beta}
            \int\prod_{(k,\kappa),(a,\alpha)} dR_{k,a}^{\kappa,\alpha}
            \int\prod_{a} dF_a \nonumber\\
 & &\times \exp\left\{ \frac{N}{2} \ln\det \left(
           {\bay{ll} {\bf P} & {\bf R}\\{\bf R}^t & {\bf Q}\eay} \right)
               + i\gamma N(1-2\delta)\sum_a F_a \right\}
   \nonumber\\
 & & \times \left[ \int_0^{\infty}\prod_{(k,\kappa)} d\rho_k^{\kappa}
                   \int\prod_{(k,\kappa)} \frac{dy_k^{\kappa}}{2\pi}
                   \int_0^{\infty}\prod_{(a,\alpha)} d\lambda_a^{\alpha}
                   \int\prod_{(a,\alpha)} \frac{dx_a^{\alpha}}{2\pi}
                   \sum_{s_k,\sigma_a} \right.\nonumber\\
 & & \;\;\;\;\;\;\times\exp\left\{ 
         -\frac{1}{2}\sum_{a,\alpha,b,\beta}Q_{a,b}^{\alpha,\beta}
x_a^{\alpha}x_b^{\beta}
         -\sum_{k,\kappa,a,\alpha} R_{k,a}^{\kappa,\alpha}y_k^{\kappa}
x_a^{\alpha}
         -\frac{1}{2}\sum_{k,\kappa,l,\lambda}P_{k,l}^{\kappa,\lambda}
y_k^{\kappa}y_l^{\lambda}
         \right.\nonumber\\
 & & \left.\left. \;\;\;\;\; \;\;\;\;\;\;\;\;\;\;\;\;\;\;
          +\;i\sum_{a,\alpha}\sigma_a\lambda_a^{\alpha}x_a^{\alpha}
          +i\sum_{k,\kappa}s_k\rho_k^{\kappa}y_k^{\kappa}
          +i\sum_a F_a \sigma_a s_1  \right\}\right]^{\gamma N}\;,
\end{eqnarray}
where $F_a$ was introduced to fix the overlap of $\bolds_1$ and 
$\boldsigma_a$ to $1-2\delta$.

The determinant can be represented by a Gaussian integral having the 
same exponent like the
quadratic part of the second exponent in (\ref{Z_mn}).
By transforming the integration variable according to
\beq
\label{factorize}
y_k^{\kappa} \mapsto y_k^{\kappa} + \sum_{l,\lambda,a,\alpha} 
({\bf P}^{-1})_{k,l}^{\kappa,\lambda}
                     R_{l,a}^{\lambda,\alpha} x_a^{\alpha}
\eeq
we obtain
\beq
\label{detdecompose}
 \ln\det \left( {\bay{ll} {\bf P} & {\bf R}\\{\bf R}^t & {\bf Q}\eay} \right)
= \ln\det{\bf P} +  \ln\det({\bf Q}-{\bf A})
\eeq
with 
$A_{a,b}^{\alpha,\beta} = \sum_{k,\kappa,l,\lambda} R_{k,a}^{\kappa,\alpha}$
$({\bf P}^{-1})_{k,l}^{\kappa,\lambda}  R_{l,b}^{\lambda,\beta}$. 
The same transformation can be 
made in the second exponent in (\ref{Z_mn}).
We analyse the resulting expression using the replica symmetric ansatz:
\begin{eqnarray}
\label{rs}
P_{k,l}^{\kappa,\lambda}&=&\left\{ \bay{ll} 1 & \;\;\;(k,\kappa)=(l,\lambda)\\
                                            P & \;\;\; k=l, \kappa\neq\lambda\\
                                            0 & \;\;\; k\neq l \eay \right. 
     \nonumber\\
 & & \nonumber\\
Q_{a,b}^{\alpha,\beta} & = &\left\{ \bay{ll} 1 & \;\; (a,\alpha)=(b,\beta)\\
                                             Q_1 & \;\; a=b, \alpha\neq\beta\\
                                             Q_0 & \;\; a\neq b \eay \right. \\
 & & \nonumber\\
R_{k,a}^{\kappa,\alpha} & = & \left\{ \bay{ll} R & \;\;\;\; k=1\\
                                               0 & \;\;\;\; k\neq 1\eay\right.
     \nonumber\\
 & & \nonumber\\
iF_a & = & F \;.\nonumber
\end{eqnarray}
$P$ describes the typical overlap within one $p$--cell 
of the global spectrum therefore fulfilling
the saddle--point equation (\ref{saddle1}) from sec. II. $\;Q_1$ 
gives the overlap of two arbitrary couplings from the same $q$--cell inside 
the cluster, $Q_0$ the overlap between two of these cells. 
Due to the fixed overlap of the cluster output with the output of the central
cell, the mirror symmetry $({\bf J},\boldsigma)\mapsto(-{\bf J},-\boldsigma)$
is explicitly broken, we
therefore expect $Q_0$ to be different from zero.  $R$ is the
overlap of the cluster cells with the central $p$--cell, whereas the 
overlap of a cell from the cluster with a randomly chosen $p$--cell
is again zero for symmetry reasons.

Finally we get the replica symmetric mass exponent by taking the 
$O(n)$-terms for $m=0$,
\begin{eqnarray}
\label{rsmass}
\tau(q|p,\delta)& =& -\frac{1}{\ln2}{\mbox{extr}}_{Q_{0,1},R,F} \left[
                       \frac{q-1}{2}\ln(1-Q_1) + \frac{1}{2} 
\ln(1+(q-1)Q_1-qQ_0)
                       +\frac{q}{2}\frac{Q_0-
\frac{p R^2}{1-(p-1)P}}{1+(q-1)Q_1-qQ_0}
                       +\gamma(1-2\delta)F \right.\nonumber\\
 & &  \left.\;\;\;\;\;\;\;\; +\gamma\frac{\int\frac{dc\;d\hat{c}}{2\pi} 
e^{ic\hat{c}}\int Dt
            \left[ \int_0^{\infty}\frac{d\rho}{\sqrt{2\pi(1-P)}}
                    \exp\{-\frac{1}{2}
\frac{(\rho-\sqrt{P}t)^2}{1-P}-i\hat{c}\rho\} \right]^{p}
            \int Du \ln \int Dw (e^F H_{-}^q + e^{-F} H_{+}^q)}{
            \int Dt H^{p}(\sqrt{\frac{P}{1-P}}t)}\right]
\end{eqnarray}
with
\beq
\label{hpm}
H_\pm = H\left( \pm \frac{\sqrt{Q_1-Q_0}w + 
\sqrt{Q_0-\frac{p R^2}{1-(p-1)P}}u+\frac{R}{1+(p-1)P}c
}{\sqrt{1-Q_1}}\right) \;.
\eeq
$P$ is given by (\ref{saddle1}) for $p$. Because of the 
integrals over complex--valued functions the general case is 
hard to handle numerically, and we concentrate on the 
most important cases $p=0,1$, i.e. the storage and generalization problems.

\section{Storage with errors}

This section we focus on the storage problem with noisy output data, 
i.e. the case of a central cell with $p=0$. Inserting this into (\ref{rsmass})
we can eliminate the integrals over complex--valued functions and find
\begin{eqnarray}
\label{storemass}
\tau(q|0,\delta) & = -\frac{1}{\ln2}{\mbox{extr}}_{Q_{0,1},F} & 
                        \left[\frac{q-1}{2}\ln(1-Q_1) + \frac{1}{2} 
\ln(1+(q-1)Q_1-qQ_0)
                       +\frac{q}{2}\frac{Q_0}{1+(q-1)Q_1-qQ_0}
                       +\gamma(1-2\delta)F \right.\nonumber\\
 & & \left. \;\;\; + \gamma \int Du \ln \int 
Dw (e^F H_{-}^q + e^{-F} H_{+}^q)\right]
\end{eqnarray}
where $H_\pm$ simplifies to
\beq
\label{hpmstor}
H_\pm = H\left( \pm \frac{\sqrt{Q_1-Q_0}w + \sqrt{Q_0}u}{\sqrt{1-Q_1}}\right) 
\;.
\eeq
The dependences on $R$ and $P$ vanish, leading to only three saddle point 
equations for the order parameters $Q_0,Q_1,$ and $F$:
\begin{eqnarray}
\label{storesaddle}
0 & = & 1 - 2\delta - \int Du \frac{\int Dw (e^F H_{-}^q - e^{-F} H_{+}^q) 
               }{\int Dw (e^F H_{-}^q + e^{-F} H_{+}^q)} \nonumber\\
0 & = & \frac{Q_0}{(1+(q-1)Q_1-qQ_0)^2} - \frac{\gamma}{2\pi(1-Q_1)}\int Du
        \left[ \frac{\int Dw (e^F H_{-}^{q-1} - e^{-F} H_{+}^{q-1}) 
               \exp\{-\frac{(\sqrt{Q_1-Q_0}w+\sqrt{Q_0}u)^2}{2(1-Q_1)}\}
                   }{\int Dw (e^F H_{-}^q + e^{-F} H_{+}^q)}
             \right]^2 \\
0 & = &  \frac{Q_1 -Q_0}{1+(q-1)Q_1-qQ_0} + 
\frac{Q_0(1-Q_1)}{(1+(q-1)Q_1-qQ_0)^2} -
              \frac{\gamma}{2\pi}\int Du
              \frac{\int Dw (e^F H_{-}^{q-2} + e^{-F} H_{+}^{q-2}) 
               \exp\{-\frac{(\sqrt{Q_1-Q_0}w+\sqrt{Q_0}u)^2}{1-Q_1}\}
                   }{\int Dw (e^F H_{-}^q + e^{-F} H_{+}^q)}
            \nonumber
\end{eqnarray}

Before solving these equations numerically, we discuss some intuitively
clear and also analytically tractable limiting cases.
For $\delta=0.5$ half the output bits are flipped and there is no remaining 
correlation between the original output sequence $\bolds$ and the sequence 
$\boldsigma$ to be learned. Up to terms irrelevant in the limit of large $N$ 
we obtain at most ${\gamma N \choose 0.5\gamma N}\simeq 2^{\gamma N}$ 
possible cells, the spectrum equals hence the global one described in Sec. II.
From the first saddle point equation we calculate $F=0$, from the
second follows $Q_0=0$. The third equation together with (\ref{storemass})  
confirm our expectation.

For $\delta=0$ both sequences $\bolds$ and $\boldsigma$ coincide up to
a non--extensive fraction of bits. The cluster thus shrinks towards its 
central cell, which has the Gardner volume. The cluster spectrum shrinks
to a single point at $\alpha_0$ (as defined in Sec. II) and $f=0$. 
In the saddle point equations (\ref{storesaddle}) we find
this result for $F\to -\infty$ and $Q_0=Q_1=P(p=0)$ fulfilling 
(\ref{saddle1}) with $p=0$.

For $q=0$ we obtain for every fixed $\delta$ the storage problem with 
an output sequence produced by flipping $\delta\gamma N$ bits randomly chosen
from a randomly drawn sequence of length $\gamma N$. The resulting output
sequence $\boldsigma$ is consequently also a random sequence of 
independent and unbiased bits. The learning problem is obviously equivalent 
to the standard Gardner problem. This is confirmed by
$\alpha(q=0|0,\delta) = \alpha_0, \forall \delta$, whereas the total number of
these cells is ${\gamma N \choose \delta\gamma N}$ resulting 
in $f(0|0,\delta)=-\gamma(\delta\log_2\delta -(1-\delta)\log_2(1-\delta))$
in the thermodynamic limit.

\begin{figure}[htb]
 \epsfysize=10cm
      \epsffile{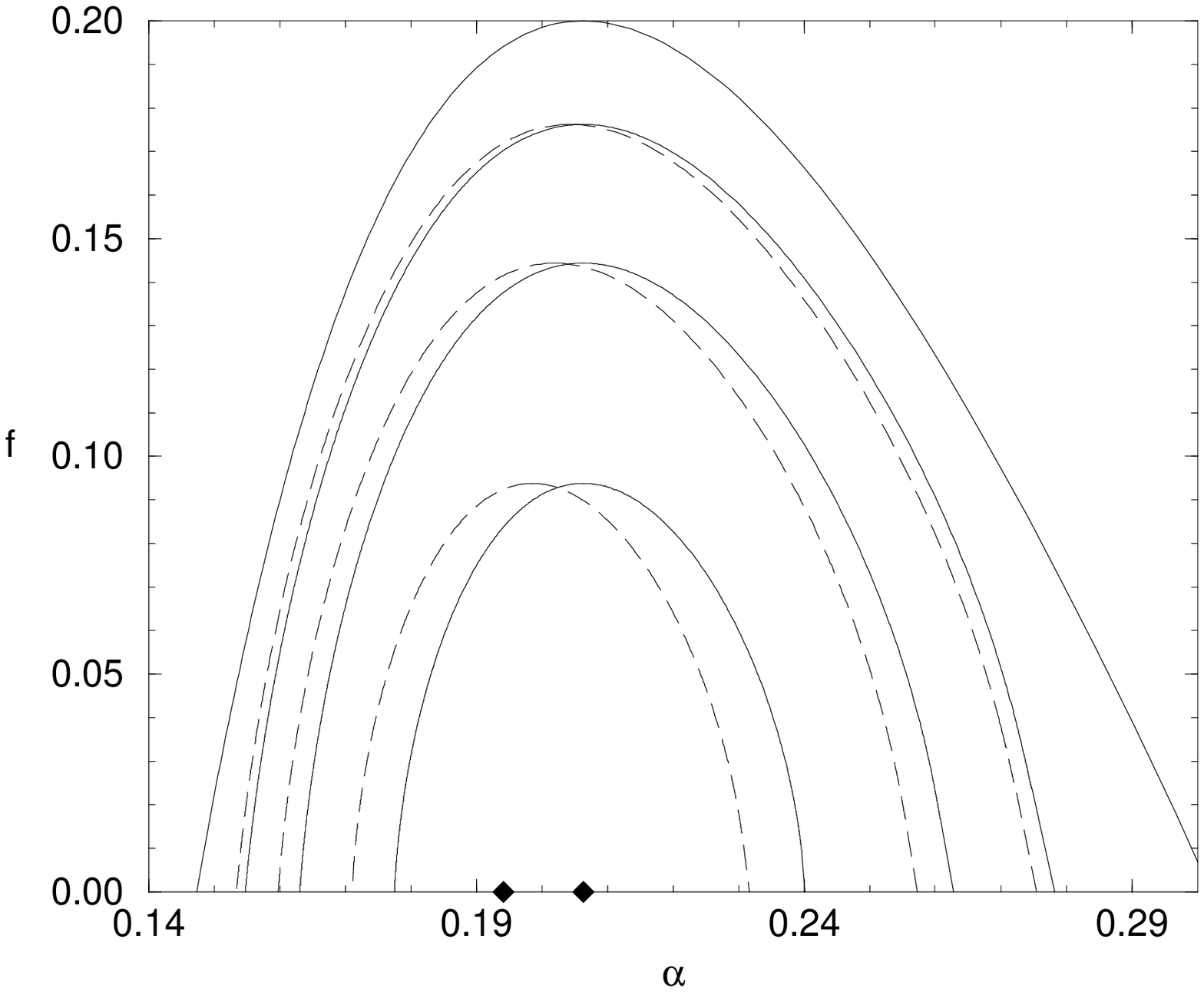}
\caption
{Multifractal spectrum $f(\alpha)$ of clusters around $p$--cells 
with $p=0,1$ (solid/dashed
lines) for $\gamma=0.2$ and $\delta=0.5,0.3,0.2,0.1$ (from top). 
The diamonds mark the crowding
indices of the central cells, they coincide with the spectra for $\delta=0$. }
\end{figure}

The rest of the spectrum has to be analyzed numerically, 
a typical set of $f(\alpha)$--curves is shown in fig. 2. The most interesting 
point is -- besides $q=0$ as discussed above --
given by $q=1$. The total volume of the cluster is given by its crowding index
$\alpha_{cl}(\delta)=\alpha(q=1|0,\delta)-f(\alpha(q=1|0,\delta))$. By 
calculating the storage capacity $\gamma_c(\delta)$ for fixed error 
rate $\delta$ from the divergence of $\alpha_{cl}$ we reproduce the replica 
symmetric results of \cite{gardner} which Gardner and Derrida calculated
for the minimal error rate above $\gamma=2$. So at least at that point, 
replica symmetry breaking effects should be taken into account in the
ansatz for the cluster overlap ${\bf Q}$. However, due to 
the complexity of even the replica symmetric calculation we refrain from
doing this. 

We still have to remark that the continuation of the mass exponent to
negative $q$ is somewhat subtle. This can be expected already by considering
the definition (\ref{pf}) of the restricted partition function 
$Z(\bolds,\delta)$. Whenever there are empty cells, (\ref{pf}) diverges
for every $q<0$, leading to $\tau(q<0|p,\delta)=-\infty$ because of the
average over all input realizations in (\ref{massexp}).
Without any change of the results for positive $q$
we can regularize $\tau$ by summing only over those $\boldsigma$ having
a non vanishing relative cell volume $P(\boldsigma)$ -- describing the 
well--defined multifractal spectrum also for $q<0$ via a 
Legendre transformation.

We consider now the last integral in (\ref{storemass}) in the case
of negative $q$. Because of 
$H(w)\propto\exp(-w^2/2)/\sqrt{2\pi}w$ for large $w$ we get an asymptotic 
exponential part of the last integrand which is proportional to 
$exp(-\Delta w^2/2+O(w))$ with
\begin{equation}
  \label{Delta}
  \Delta=\frac{1+(q-1)Q_1-qQ_0}{1-Q_1}\;.
\end{equation}
The integral consequently diverges for $\Delta\geq0$, i.e. for every 
$0\leq Q_0\leq 1$ at 
$Q_1=(1-qQ_0)/(1-q)$, and the global minimum in (\ref{storemass}) 
with respect to $Q_1$ is no longer
given by the saddle point equations (\ref{storesaddle}). Due to this the
mass exponent would be expected to diverge to $-\infty$ for every $q<0$. 
On the other hand,
the continuation of the saddle point equations (\ref{storesaddle}) to $q<0$
gives smooth results for the mass exponent and the multifractal spectrum.
We expect it therefore to describe the correct regularization of the
partition function at least within the replica symmetric approximation.

\section{Generalization with errors}

In this section we treat the question of generalizing noisy output data.
As mentioned in Sec. II, this problem corresponds to taking $p=1$. 
Also in this case the complex--valued integrals can be evaluated analytically.
The mass exponent is given by
\begin{eqnarray}
\label{genmass}
\tau(q|1,\delta) & = -\frac{1}{\ln2}{\mbox{extr}}_{Q_{0,1},R,F} & 
                        \left[\frac{q-1}{2}\ln(1-Q_1) + \frac{1}{2} 
\ln(1+(q-1)Q_1-qQ_0)
                       +\frac{q}{2}\frac{Q_0-R^2}{1+(q-1)Q_1-qQ_0}
                       +\gamma(1-2\delta)F \right.\nonumber\\
 & & \left. \;\;\; + 2\gamma \int Du H\left(\frac{uR}{\sqrt{Q_0-R^2}}\right) 
                \ln \int Dw (e^F H_{-}^q + e^{-F} H_{+}^q)\right]
\end{eqnarray}
with
\beq
\label{hpmgen}
H_\pm = H\left( \pm \frac{\sqrt{Q_1-Q_0}w + \sqrt{Q_0}u}{\sqrt{1-Q_1}}\right) 
\;.
\eeq
Again, the dependence on $P$ vanishes whereas $R$ remains an order 
parameter to be optimized. We obtain four saddle point equations
which determine $Q_0, Q_1, R$, and $F$:
\begin{eqnarray}
0 & =&  1 - 2\delta - 2\int Du\;H\left(\frac{uR}{\sqrt{Q_0-R^2}}\right) 
         \frac{\int Dw (e^F H_{-}^q - e^{-F} H_{+}^q) 
               }{\int Dw (e^F H_{-}^q + e^{-F} H_{+}^q)} \nonumber
\end{eqnarray}
\begin{eqnarray}
\label{gensaddle}
0 & = & \frac{Q_0-R^2}{(1+(q-1)Q_1-qQ_0)^2}- \frac{\gamma}{\pi(1-Q_1)}\int Du
          \;H\left(\frac{uR}{\sqrt{Q_0-R^2}}\right)\nonumber\\
  &   &  \times
        \left[ \frac{\int Dw (e^F H_{-}^{q-1} - e^{-F} H_{+}^{q-1}) 
               \exp\{-\frac{(\sqrt{Q_1-Q_0}w+\sqrt{Q_0}u)^2}{2(1-Q_1)}\}
                   }{\int Dw (e^F H_{-}^q + e^{-F} H_{+}^q)}
             \right]^2
\end{eqnarray}
\begin{eqnarray}
0 & = &  \frac{Q_1 -Q_0}{1+(q-1)Q_1-qQ_0} + 
\frac{(Q_0-R^2)(1-Q_1)}{(1+(q-1)Q_1-qQ_0)^2}
          \nonumber\\
  &   &  -\frac{\gamma}{\pi}\int Du \; H\left(\frac{uR}{\sqrt{Q_0-R^2}}\right)
              \frac{\int Dw (e^F H_{-}^{q-2} + e^{-F} H_{+}^{q-2}) 
               \exp\{-\frac{(\sqrt{Q_1-Q_0}w+\sqrt{Q_0}u)^2}{1-Q_1}\}
                   }{\int Dw (e^F H_{-}^q + e^{-F} H_{+}^q)}
            \nonumber 
\end{eqnarray}
\begin{eqnarray}
0 &=& \frac{qR}{1+(q-1)Q_1-qQ_0} + \frac{\gamma}{\pi}\int du 
                    \;e^{-\frac{Q_0 u^2}{2(Q_0-R^2)}}
\frac{Q_0 u}{(Q_0-R^2)^{3/2}}
                   \; \ln \int Dw (e^F H_{-}^q + e^{-F} H_{+}^q)\;. \nonumber
\end{eqnarray}
Several intuitively clear limiting cases can be discussed analytically. 
As argued in the previous section, for $\delta=0.5$ we recover the full 
spectrum with order parameters $F=0, R=0, Q_0=0$ and $Q_1=P(q)$. $R$ is the
overlap of the central cell with the cells of the cluster. Its value is found
to be zero for all $q$ indicating that all types of cluster cells are 
orthogonal to the teacher vector, their volumes are dominated by the part 
lying on the $(N-1)$--dimensional ``equator''.
The learning of a vector obtained by flipping half of the teacher's outputs 
is obviously equivalent to the storage problem of a random output sequence. 
The student is not able to get any information about the teachers rule.

For $\delta\rightarrow 0$ the cell cluster shrinks towards the central 
cell, which is the version space of the corresponding  noise--free 
generalization problem. $F$ diverges to $-\infty$
whereas the other three order parameters coincide asymptotically, 
$Q_0=Q_1=R=P(p=1)$.
The crowding index takes only the value $\alpha_1$, $f(q|1,0)$ is 
found to be zero.
The equivalence of this solution to earlier results of \cite{GT} 
was already discussed in \cite{EW,WE}.

For general $\delta$ the analysis has to be done numerically. 
In fig. 1 we show a
representative set of spectra for several values of $\delta$. For growing
error rate not only the number of cells in the cluster increases, but also the
range of different existent cell sizes.

For $q=0$ we obtain a student who perfectly learns an output sequence 
generated by
a teacher, but flipped with rate $\delta$. This corresponds to the case of
output noise analyzed in \cite{GT,OpKi}. The cell sizes go for $0<\delta<0.5$
from $\alpha_1$ to $\alpha_0$, thus interpolating between the noiseless
learning from examples and the storage problem for random input--output
relations. This interpretation
gives also a sense to the part of the global cell spectrum for inverse
temperatures between zero and one, at least a proper subset of these can
be understood as generalization tasks including noisy output data.
Of course, for $\delta>0$ this
task is not learnable for large loading ratios $\gamma$. This observation
leads directly to a storage capacity $\gamma_c(\delta)$ going monotonously
from $\gamma_c(0)=\infty$ to the Gardner value $\gamma_c(\delta=0)=2$. 
For every 
$\delta$ the overlap $R$ between teacher and student is a monotonously
increasing function of the loading ratio. Its maximal value is reached
at $R_{max}(\delta)=R(\gamma_c(\delta))$ which remains strictly smaller
then one for every $\delta\neq 0$.

\begin{figure}[htb]
 \epsfysize=10cm
      \epsffile{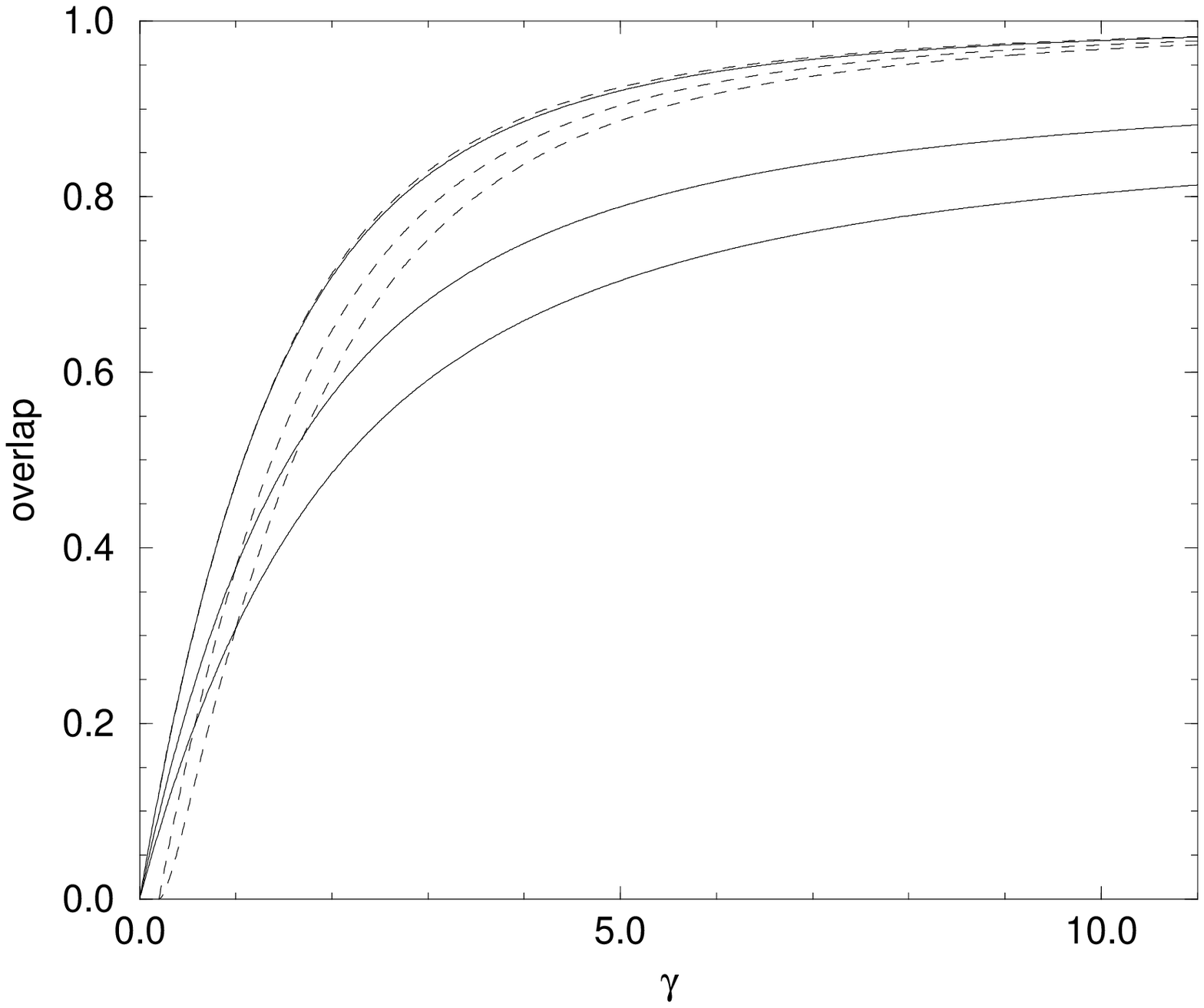}
\caption
{Overlaps $Q_1, R, Q_0$ (from top to bottom) for the generalization 
problem with a student making up to $0.1\gamma N$ (full lines) and
$0.1N$ (dashed lines) errors}
\end{figure}

Another problem can be analyzed in the spectrum at $q=1$. The total volume
of the cell cluster surrounding a cell of size $\alpha_1$ is given
by its total crowding index which can again be calculated from 
$\alpha_{cl}(\delta) = \alpha(q=1|1,\delta) - f( \alpha(q=1|1,\delta))$.
This learning task corresponds to a lazy student being satisfied with any
output having at most $\delta\gamma N$ errors compared with the sequence
of examples presented by the teacher, cf. \cite{GT}. The student can achieve 
this for every value of $\gamma$, an upper threshold for the loading ratio 
does not exist. 
As illustrated by the full lines in fig. 3, in the case of a fixed error rate 
$\delta>0$ the overlap $R$ between teacher and student does not go to one, and
the generalization error $\varepsilon = \frac{1}{\pi}\arccos R$ does not
tend to zero for increasing loading ratio $\gamma$. The cell volume of every
special output realization shrinks to zero, $Q_1\to1$, but this is
compensated by a cell number exponentially growing with $\gamma N$.
Thus, the resulting total cluster volume does not vanish, $Q_0<1$.
If we fix instead the total number of errors, the number of possible
representations does not depend on $\gamma$ either. The vanishing version space
volume of every particular output sequence thus results in a vanishing 
total volume, 
leading to a vanishing generalization error in the limit of large loadings 
$\gamma$, cf. the dashed lines in figure 3. In both cases, the information 
gain \cite{OpKi}
$\partial \alpha_{cl} / \partial \gamma$ goes from values of order
one (halving the cell with every new pattern) for small $\gamma$ to
zero for $\gamma \to \infty$.

The inclusion simultaneous noise for teacher and student requires the 
introduction of 6 different replications resulting in an even more complex
structure of the order parameter equations. Therefore we refrain from doing 
it.

\section{On the spatial cell distribution}

From the spectrum of the internal representations in a cluster we can also
get some information on the spatial distribution of the cells. If the latter
were totally random, we would not expect any dependence of the
internal cluster spectrum on the central cell, i.e. on $p$. In this case,
cells of all possible sizes should be contained in the cluster. From fig. 1,
where the spectra are plotted for $p=0,1$, we can deduce that the
distribution has some structure. Reducing $\delta$ from $0.5$ does not
only decrease the number of cells, but also of the range of different
cells. Both very large as well as very small cells are excluded.

In the neighborhood of $\delta=0$ the spectrum is
concentrated in a small interval around the crowding index $\alpha_p$
of the central cell. This means that every cell is
surrounded by cells having almost the same size leading to some
kind of clustering of cells of nearly equal size. So there appear in the 
neighborhood of very large cells no very small cells and vice versa. Of 
course, due to the symmetry of the probability distributions
for the input patterns, these ``clusters'' of nearly equally sized
cells are isotropically located in coupling space.

\section{Summary}

In the present paper we have analyzed the internal structure of cell clusters
having a given output overlap with a certain central cell. The 
calculation of the internal multifractal spectrum of such clusters
allowed us to discuss various storage and generalization problems of noisy
output data within one single unified approach. The analysis included
both the case of a lazy student which is satisfied whenever he achieves
some maximal error rate, as well as the case of an absent--minded
teacher offering incorrect data to his student. In the global cell 
spectrum of the whole coupling space it was not possible to give an
interpretation to cells of crowding indices in--between
the Gardner value $\alpha_0$ and the generalization value $\alpha_1$. As a
result of the present approach we are able to understand at least a proper 
subset of these cells as related to generalization tasks with output noise. 
Additionally we have shown that every cell is surrounded by cells
having nearly the same size. The range of realized sizes is increasing
with decreasing overlap of the output sequences labeling the cells.

We are aware of the fact that the multifractal approach is plagued by 
the existence of replica symmetry breaking, but due to the
technical difficulties of a calculation which includes four different
kinds of replicas we restricted our analysis to the replica symmetric
ansatz. The inclusion of replica symmetry breaking effects would 
surely change some of the calculated quantities, but the qualitative
picture would probably remain unchanged.

{\bf Acknowledgments:} Many thanks to A. Engel and J. Berg for
illuminating discussions and for careful reading the manuscript.

\end{document}